\documentclass{elsart}
\usepackage[dvips,final]{graphicx}
\usepackage{epsfig}
\usepackage{subfigure}
\usepackage{epic}
\usepackage{eepic}
\usepackage{amssymb,amsmath}
\usepackage{times}

\begin{document}

\hyphenation{di-men-sio-nal}
\newcommand{\fixme}[1]{ { \bf{ ***FIXME: #1 }} }

\runauthor{J.~Harting, M.J.~Harvey, J.~Chin and P.V.~Coveney}
\begin{frontmatter}
\title{Detection and tracking of defects in the gyroid mesophase}
\author[ica]{Jens Harting},
\author[ucl]{Matthew J. Harvey},
\author[ucl]{Jonathan Chin},
\author[ucl]{Peter V. Coveney}

\address[ica]{Institute for Computational Physics, University of Stuttgart, Pfaffenwaldring 27, D-70569 Stuttgart, Germany}
\address[ucl]{Centre for Computational Science, Christopher Ingold Laboratories, University College London, 20 Gordon Street, London WC1H 0AJ, UK}
\begin{abstract}
Certain systems, such as amphiphile solutions or diblock copolymer melts,
may assemble into structures called ``mesophases'', with properties
intermediate between those of a solid and a liquid. These mesophases can
be of very regular structure, but may contain defects and grain
boundaries. Different visualization techniques such as volume rendering or
isosurfacing of fluid density distributions allow the human eye to detect
and track defects in liquid crystals because humans are easily capable of
finding imperfections in repetitive spatial structures.  However, manual
data analysis becomes too time consuming and algorithmic approaches are
needed when there are large amounts of data. We present and compare two
different approaches we have developed to study defects in gyroid
mesophases of amphiphilic ternary fluids. While the first method is based
on a pattern recognition algorithm, the second uses the particular
structural properties of gyroid mesophases to detect defects.
\end{abstract}
\end{frontmatter}

\section{Introduction}
Molecules in an ordinary liquid will usually have random orientations.
There is a certain interesting class of liquids, called liquid crystals,
in which this is not the case: the molecules exhibit a tendency to align
or order.  Liquid crystals have some of the properties of crystals,
since they exhibit long-range ordering and strong anisotropy, but retain
the ability to flow, unlike ordinary crystals in which the molecules are
locked onto well-defined positions on a lattice.

Amphiphiles are one class of molecule which may produce liquid
crystalline behaviour. These are molecules constructed from two parts,
usually a water-loving head and a long tail which is attracted to oil.
In a mixture of oil, water, and amphiphile, amphiphile molecules are
attracted to the interface between the oil and water to minimize free
energy, hence they are often termed ``surface active agents'', or
surfactants. Their precise behaviour is strongly dependent on
concentration, so they are termed ``lyotropic'' liquid crystals.

A random mixture of oil, water, and surfactant molecules will often
spontaneously arrange itself into separate regions of oil and water,
separated by a layer of surfactant at the interface, forming structures
called surfactant mesophases. These structures can be very complex, and
their geometry depends strongly upon the relative proportions of
different molecules in the mixture, and the details of how they
interact. Such structures can also be produced in binary systems, such
as water-lipid mixtures, and often occur in biological
systems\cite{bib:seddon-templer,bib:seddon-templer-2,bib:czeslik-winter}.

Surfactant mesophases will often form minimal surfaces: for any boundary
drawn on the surface, the surface lying inside the boundary will have the
minimal possible surface area, and the surface will also have zero mean
curvature.  Triply periodic minimal surfaces\cite{bib:karcher-polthier}
which extend through space with cubic symmetry\cite{bib:lord-mackay} have
been found in many systems, such as lipid-water
mixtures\cite{bib:seddon-templer}, diblock
copolymers\cite{bib:shefelbine-vigild-etal}, and in many biological
systems\cite{bib:landh}.

The gyroid, also known as the ``G surface'', is a particular minimal
surface, discovered by Schoen\cite{bib:schoen}. It is embedded
(has no self-intersections), triply periodic (repeats in $x$, $y$, and
$z$ directions), and is the only known such surface with triple
junctions (Figure \ref{fig:doublegyroid-wishbones}). A numerical study
of the gyroid was made by
Gro{\ss}e-Brauckmann\cite{bib:grosse-brauckmann}; while an analytical
description was found by Gandy and Klinowski\cite{bib:gandy-klinowski}
a very close approximation of a gyroid is the surface $ \cos x \sin
y + \cos y \sin z + \cos z \sin x = 0 $.  It is possible to transform the
gyroid into two other well-known surfaces, the P and D surfaces, through
the single parameter known as the Bonnet Angle\cite{bib:fogden-hyde}.
The gyroid has symmetry group $\mathrm{Ia\bar{3}d}$; the unit cell
consists of 96 copies of a fundamental surface patch, related through
the symmetry operations\cite{bib:gandy-klinowski} of this space group.

The gyroid surface divides space into two interpenetrating regions, or
labyrinths. In the case of gyroids formed from a mixture of oil, water,
and surfactant, one labyrinth contains mostly oil, the other mostly
water, and the gyroidic boundary surface between the two labyrinths is
populated with surfactant, as shown in Figure \ref{fig:dirsurslice} and
\ref{fig:doublegyroid-wishbones}.


\newcommand{\jonfigsize}{4.5cm}
\begin{figure}

\begin{center}

\mbox{
        \subfigure[]{
                \includegraphics[width=\jonfigsize,height=\jonfigsize]{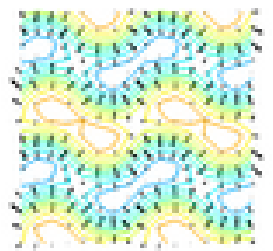}
                \label{fig:dirsurslice}
        } \quad
        \subfigure[]{
                \includegraphics[width=\jonfigsize,height=\jonfigsize]{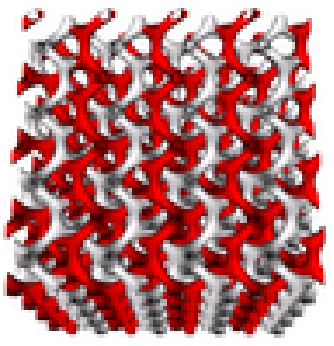}
                \label{fig:doublegyroid-wishbones}
        }
}

\mbox{
        \subfigure[]{
                \includegraphics[width=\jonfigsize,height=\jonfigsize]{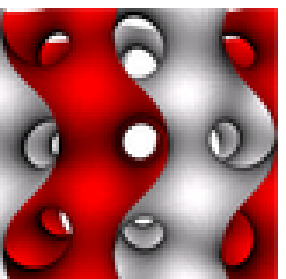}
                \label{fig:contrarotation}
        } \quad
        \subfigure[]{
                \includegraphics[width=\jonfigsize,height=\jonfigsize]{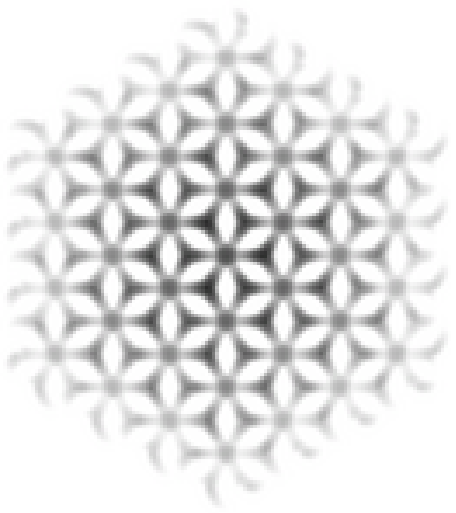}
                \label{fig:gyroid-ortho-volume}
        }
}

\end{center}

\caption{
Views of the gyroid mesophase.  \ref{fig:dirsurslice} 2D slice from a
small ($16^3$) simulation of a gyroid mesophase.  Contours show
composition, varying from pure oil to pure water; arrows show the
surfactant orientation. It can be seen clearly that the surfactant sits
at the interface, with the head groups pointing towards the water
component.
\ref{fig:doublegyroid-wishbones} Structure of the two labyrinths
enclosed by a gyroid minimal surface, showing the characteristic triple
junctions.
\ref{fig:contrarotation} Channels running in the (100) direction of a
gyroid surface: note how adjacent channels rotate in opposite senses. 
\ref{fig:gyroid-ortho-volume} Parallel-projection volume rendering
of a gyroid, looking in the (111) direction to show the distinctive
``wagon-wheel'' appearance.
}

\end{figure}


Channels run through the gyroid labyrinths in the (100) and (111)
directions; passages emerge perpendicular to any given channel as it is
traversed, the direction at which they do so gyrating down the channel,
giving rise to the ``gyroid'' name\cite{bib:grosse-brauckmann}.

The labyrinths are chiral, so that the channels of one labyrinth
gyrate in the opposite sense to the channels of the other, as seen in
Figure \ref{fig:contrarotation}. Looking down the (111) direction of a
gyroid shows a distinctive ``wagon wheel'' pattern (Figure
\ref{fig:gyroid-ortho-volume}), which has been observed experimentally
in transmission electron micrographs of gyroid
phases\cite{bib:shefelbine-vigild-etal}.

As a Platonic or mathematical abstraction, the gyroid consists of
perfect copies of the unit cell, repeating on a Bravais lattice
extending through space.  This is not the case for gyroid structures in
the real world: various effects may give rise to regions where the
structure deviates from a gyroid.  Such deviations are called {\em
defects}. 

During the gyroid self-assembly process, several small, separated
gyroid-phase regions or domains may start to form, and then grow. Since
the domains evolve independently, the lattices describing them may not
be identical, and can differ in orientation, position, or unit cell
size.  The interface between the domains will not be gyroidal:
therefore, grain-boundary defects arise between gyroid domains. Inside a
domain, there may be dislocations, or line defects, corresponding to the
termination of a plane of unit cells; there may also be localised
non-gyroid regions, corresponding to defects due to contamination or
inhomogeneities in the initial conditions.

\begin{figure}
\begin{center}
\includegraphics[width=5cm,height=5cm]{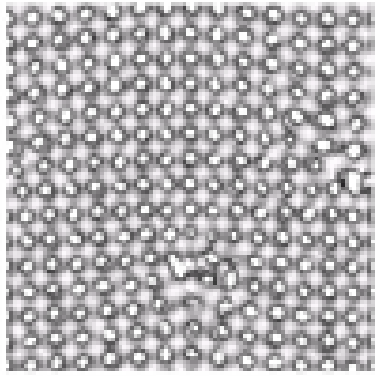}
\caption{
Dislocation in a simulated gyroid mesophase
\label{fig:dislocation}
}
\end{center}
\end{figure}

While equilibrium gyroid mesophases and their defects are observed
experimentally, it is desirable to formulate a theoretical or
computational model to better understand how and why they form and how
they evolve.

Much effort has been invested in theoretical and computational modelling of
liquid crystals. Nematic liquid crystals have been modelled using the
Leslie-Erickson formulation \cite{bib:degennes-prost}; Monte
Carlo\cite{bib:frenkel-eppenga,bib:allen} and Molecular
Dynamics\cite{bib:allen-warren} simulations have provided some insight, but
reaching regimes where hydrodynamic effects are significant is currently
computationally unfeasible with these techniques.

Lyotropic liquid crystals, of interest in this paper, have also been
studied extensively\cite{bib:gompper-schick} through techniques such as
free-energy methods\cite{bib:ciach-holyst,bib:holyst-oswald} and through
consideration of the interface between the lipid and water
phases\cite{bib:schwartz-gompper}; again, most treatments have been
limited to examination of the equilibrium state and its stability.

There have been recent attempts to take advantage of the lattice Boltzmann
method for hydrodynamics\cite{bib:succi,bib:qian-dhumieres-lallemand}, and
modify it to take account of liquid-crystalline behaviour.  Lattice Boltzmann
is a discrete-time and discrete-space algorithm; since only nearest-neighbour
interactions take place between vertices on the simulation lattice, it is
extremely fast, and extremely scalable on parallel computer
hardware\cite{bib:love-nekovee-coveney-chin-gonzalez-martin}. There are several
such schemes, some of which\cite{bib:care-halliday-good} are based around the
Leslie-Ericksen model for nematic liquid crystals, and others which use a
free-energy-based approach for
nematic\cite{bib:swift-orlandini-osborn-yeomans,bib:orlandini-swift-yeomans,bib:swift-osborn-yeomans,bib:denniston-marenduzzo-orlandi-yeomans,bib:denniston-orlandini-yeomans,bib:denniston-orlandini-yeomans2}
and lyotropic\cite{bib:theissen-gompper-kroll,bib:lamura-gonnella-yeomans}
phases.

We used another kind of lattice Boltzmann model, which employs a ``bottom-up''
approach to model interactions between
particles\cite{bib:shan-chen,bib:martys-douglas} including
surfactants\cite{bib:chen-boghosian-coveney}, postulating a form of
interparticle potential rather than using a free-energy-based technique.
Briefly, the single-particle distribution function $f_i^{\sigma}({\bf x})$ for
species $\sigma$ with velocity ${\bf c}_i$ is discretized onto lattice points
$\bf x$, and then evolved according to the lattice BGK
equation\cite{bib:qian-dhumieres-lallemand}. Three species, called red, blue,
and green, are used, corresponding to oil, water, and surfactant.  Immiscible
fluids are modelled using an interparticle interaction force, controlled by a
coupling constant $g_{cc}$: this force is calculated from the gradient of the
order parameter, or ``colour field'', defined as the difference $\phi(\bf
x)=\rho^r (\bf x) - \rho^b (\bf x)$ between red and blue fluid densities. The
mean surfactant director field ${\bf d}({\bf x})$ is also tracked on the
lattice. A point-like surfactant molecule is modelled as being constructed from two
different immiscible fluid molecules joined together, and therefore subject to
dipole-like interactions with the other fluids, controlled by a coupling
constant $g_{cs}$. Finally, interactions between surfactant molecules are
controlled by a constant $g_{ss}$. It was recently
shown\cite{bib:gonzalez-coveney,bib:gonzalez-coveney-2} through simulations
with the LB3D parallel lattice Boltzmann code that certain mixtures of specific
composition (specified by the initial densities $f_r$,$f_b$,$f_g$ of oil,
water, and surfactant) would spontaneously assemble from a randomized,
disordered initial condition, into a gyroid mesophase whose lattice parameter
is around $8$--$9$ simulation lattice sites.  Similar cubic phases have been
observed experimentally with lattice parameter of order $50\mathrm{nm}$, in
polymer blends\cite{bib:shefelbine-vigild-etal} and biological
systems\cite{bib:landh}.

On a sufficiently small lattice, the gyroid may evolve to perfectly fill the
simulated region, without defects. As the lattice size grows, it becomes more
probable that multiple gyroid domains will emerge independently, so that grain
boundary defects are more likely to appear and the time required for localized
defects to diffuse across the lattice increases making it more likely that
defects will persist.  Therefore, examination of the defect behaviour of
surfactant mesophases requires the simulation of very large systems. This was
achieved as part of the TeraGyroid project, where systems on lattice sizes of
up to $1024^3$ were simulated by linking together multiple
geographically-distributed supercomputing resources to form a computational
Grid\cite{bib:teragyroid,bib:chin-harting-jha,bib:ReG,bib:TeraGyroidWWW}.

\section{Structure factor analysis of liquid gyroid mesophases}
Simulation data from liquid crystal dynamics can be visualized using
isosurfacing or volume rendering techniques. The human eye has a
remarkable ability to easily distinguish between regions where the crystal
structure is well developed and areas where it is not. For quantitative
studies of large systems evolving over long intervals of time,
computational methods for defect detection and tracking are required.
Developing algorithms to detect and track defects is a non-trivial task,
however, since defects can occur within and between domains of varying
shapes and sizes and over a wide variety of length and time scales. 

A standard method to analyse simulation data is the calculation of the
three-dimensional structure function 
\begin{equation}
S(\mathbf{k},t)\equiv\frac{1}{V}\left|\phi^\prime_\mathbf{k}(t)\right|^2,
\end{equation} 
where $V$ is the number of cites of the lattice,
$\phi^\prime_\mathbf{k}(t)$ the Fourier transform of the fluctuations of
the order parameter $\phi^\prime\equiv\phi-\left<\phi\right>$, and
$\mathbf{k}$ is the wave vector
\cite{bib:gonzalez-nekovee-coveney,bib:gonzalez-coveney-2}.  $S(\mathbf{k},t)$
can easily be calculated, but only gives general information about the
crystal development \cite{bib:hajduk,bib:laurer,bib:gonzalez-coveney}. It
does not allow one to detect where the defects are located or how many
there are, nor does it furnish access to information about the number
of differently oriented gyroid domains. 

\begin{figure}[h]
\begin{center}
\includegraphics[height=8cm]{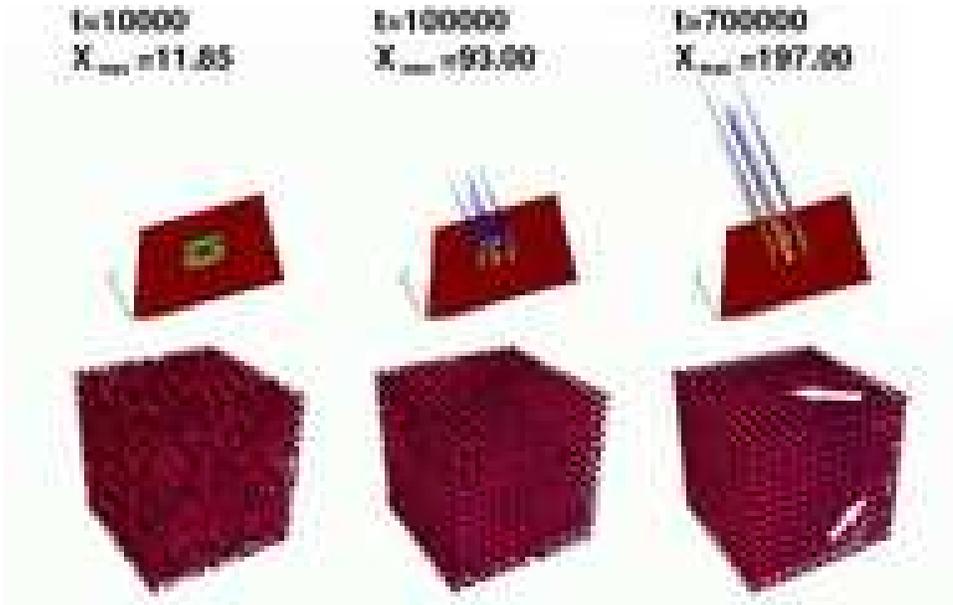}
\end{center}
\caption{Three-dimensional structure factor of the order parameter at
timesteps $t$=10000, 100000, and 700000, lattice size 128$^3$ and
simulation parameters $f_r$=$f_b$=0.7, $f_g$=0.60, $g_{ss}$=-0.0045,
$g_{cs}$=-0.006. For comparability with SAXS experimental data, we
display the total structure factor in the $x$-direction $X$=$\sum_{k_x}
S(\mathbf{k},t)$. $X_{max}$ denotes the value of the largest peak divided
by the number of lattice sites in the direction of summation (128 in this
case). The lower half of the figure shows volume rendered visualizations
of the corresponding order parameters and the white arrows are a guide for
the eye to spot some defective areas at the top surface and the right
corner at $t$=700000.}
\label{fig:3DFFT}
\end{figure}

Figure \ref{fig:3DFFT} gives an
example of the three-dimensional structure factor calculated for the order
parameter $\phi(\mathbf{x})$ at timesteps $t$=10000, 100000, and 700000.
We use periodic boundary conditions, a 128$^3$ lattice and simulate for
one million timesteps. This is more than an order
of magnitude longer than any other simulation performed before the
TeraGyroid \cite{bib:TeraGyroidWWW} project using our lattice-Boltzmann code
LB3D and took 300 wall clock hours on 128 CPUs of an IBM SP4 (namely HPCx
in Daresbury, UK). We use data from this simulation throughout the present
paper to demonstrate the properties of different defect detection and
tracking algorithms. The initial condition of the simulation is a random
mixture with maximum densities of 0.7 for the case of the immiscible
fluids and 0.6 for surfactant. The surfactant-surfactant coupling constant
is given by $g_{ss}$=-0.0045 and the coupling between surfactant and the
other fluids is determined by $g_{cs}$=-0.006. In order to obtain a visualization that is comparable
to experimentally obtained SAXS data (see for example
\cite{bib:hajduk}), we sum the structure factor in one of the cartesian
directions. The example here has been summed in the $x$-direction and
$X_{max}$ denotes the value of the largest peak normalised by the number
of lattice sites in the direction of summation (128 in this case). Results
for the $y$- and $z$-directions are similar. At $t$=10000, gyroid assembly
is just commencing, which is evident due to the eight peaks of the
structure factor which are already clearly distinguishable. $X_{max}$ is
11.85 in this case and is almost eight times bigger for $t$=100000. The
peaks of the structure function correspond to a well developed crystal
which consists of differently oriented gyroid domains with defects at the
domain boundaries. At $t$=700000, $X_{max}$ reaches 197.00 and most of the
previously existing domains have merged into a single one. Only a few
defects are left of which two can be spotted visually at the right corner
of the volume rendered visualisation and the centre of the top surface
(denoted by the white arrows).

\begin{figure}[h]
\begin{center}
\includegraphics[height=6cm]{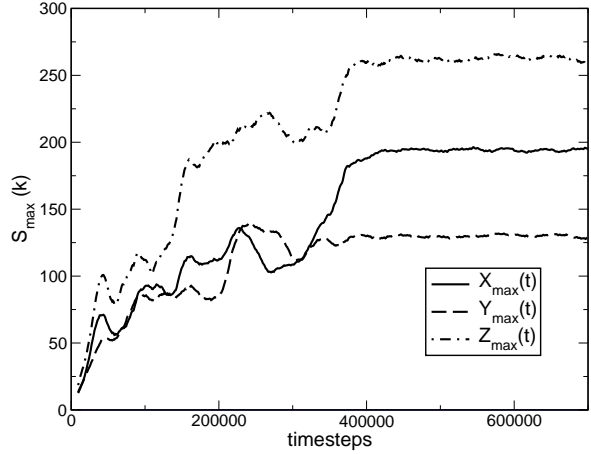}
\end{center}
\caption{Maximum value of the three othogonal projections of the
three-dimensional structure factor. For a clearer presentation, we
averaged over 20000 timesteps to obtain a single data point displayed
here.} 
\label{fig:SFMAX}
\end{figure}

We are interested here in very well developed liquid crystals with defect
domains covering only a minority of the total simulation volume. In order
to distinguish between different defects and to study their evolution in
time, we need to be able to clearly separate defect domains from regions
where we find a perfect gyroid structure. An important property of these
systems is that the variations of the structure function in time become
very small and the system reaches a state close to equilibrium. Figure
\ref{fig:SFMAX} shows the time dependence of the maxima of the structure
function in $x$-, $y$- and $z$-direction for up to 700000 timesteps. To
suppress short lived fluctuations within the fluid mixture, i.e. local
variations that spontanously form and disappear after up to a few thousand
timesteps, we average every data point over 20000 timesteps. The maximum
value of $S(\mathbf{k},t)$ shows a generally increasing behaviour in all
three cases, but fluctuates greatly for the first 320000 timesteps. Then,
$X_{max}(t)$ and $Z_{max}(t)$ show a steep increase indicating that two
major gyroid domains are merging into a single one. During this process,
defects located at the boundaries between these domains disappear. At
$t$=400000, the fluctuations present in all three plots become very small
indicating a very clean crystal with only a small number of defects. It is
easy to detect this state numerically by defining a maximum allowed
variation of the maximum values of the structure function. The remaining
part of this paper will only discuss the analysis of data obtained for
$t\ge$340000. 

Very long simulations like the one presented here can generate large
amounts of data -- especially if one measures physical quantities with a
high resolution.  Here, we measured the order parameter every 100
timesteps resulting in 10000 data files or about 78GB of data which we
need to analyse. Filtering out data that is irrelevant for studying defect
behaviour using the method described here allows us to reduce the number
of files to 6500. 
\section{Data reduction: From three dimensions to two dimensions}
A first order approach to reduce the amount of data which needs to be
analysed in detail is to project the three-dimensional system onto a
two-dimensional plane. If one volume renders the order parameter of a
gyroid unit cell using a step function which fills out all areas above an
appropriate threshold value and leaves all values below that threshold
transparent, it is possible to ``look through'' the unit cell under
various angles. Since in a perfect gyroid mesophase the individual unit
cells assemble in a very regular way, it is then possible to look through
the whole liquid crystal.  This can be implemented as a ray-tracing
algorithm:  First, select an appropriate projection direction, for example
$y$. Define a projection plane $P(x,z)$ to store the results and define a
threshold value $C$. For the gyroid mesophase, we use 66\% of the maximum
value of the order parameter, but this value depends on the system to be
analysed. For every point in the $xz$-plane, start at $y$=0 and check for
all values of $y$ if the order parameter $\phi(x,y,z)$ is smaller than
$C$. As long as that is the case, we keep $P(x,z)$=0. If $\phi(x,y,z)$ is
greater or equal $C$, we set $P(x,z)$=1, move to the next point in the
$xz$-plane and start from $y$=0 again.  

\begin{figure}[h]
\begin{center}
\includegraphics[height=7cm]{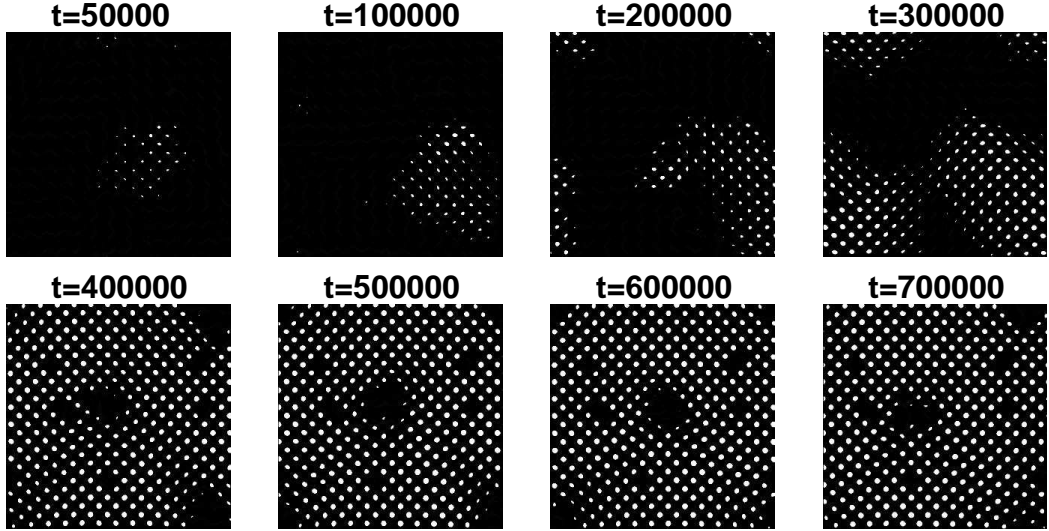}
\end{center}
\caption{Two-dimensional projection in the $y$-direction of volume rendered
three-dimensional colour fields at $t$= 50000, 100000, 200000,
300000, 400000, 500000, 600000, 700000.}
\label{fig:colourblack}
\end{figure}

Figure \ref{fig:colourblack} shows visualizations of $P(x,z)$ for timesteps
50000, 100000, 200000, 300000, 400000, 500000, 600000, 700000. Black areas
correspond to $P(x,z)$=1 and white areas to $P(x,z)$=0. For early simulation
times most of $P(x,z)$ is 1 and while the simulation evolves white spots start
to occur in the images presented in figure \ref{fig:colourblack} until a very
regular lattice-like structure with some black islands appears for $t>$400000.
Areas of regular lattice structure correspond to perfect crystal structures
along the projection direction. Black islands can be interpreted as areas where
the gyroid structure is disturbed or not existent at some point along the
projection axis.

Obviously, projecting the full system makes it impossible to retain the
three-dimensional structure of individual defects. Therefore, we apply the
projection algorithm on slabs of the dataset only. For an optimal
resolution of the defect detection, the slab thickness $l$ should be
comparable to the size of a gyroid unit cell which corresponds to eight
lattice sites in our case, resulting in 16 individual
128$\times$128$\times$8 slabs for a 128$^3$ lattice. We found
that using overlapping slabs does not improve the defect detection rate.
The positions and sizes of the defects detected in the two-dimensional
projections can be used to reconstruct three-dimensional datasets which
only include the defective areas. A defective area in the two-dimensional
datasets is mapped to a volume of thickness $l$.

In order to further improve the reliability of the detection, we repeat
the analysis for all three cartesian directions. In this way, we can 
detect gyroid cells which are deformed in one direction only. For the
reconstruction of the three-dimensional dataset, all three analysis runs
are taken into account. Additional resolution can be obtained by
distinguishing between how often and in which direction(s) a defective
volume has been detected, indicating the particular kind of defect.

The human eye is easily able to accurately detect defective areas in the
individual images of figure \ref{fig:colourblack}. In the following
sections we will present two possible approaches which try to transfer
this remarkable ability to a well defined algorithm that can be
implemented on a computer. The first approach is based on a generic
pattern recognition algorithm and should work with all liquid crystals
that form a regular pattern, while the second has been developed with our
particular problem in mind and is not known to work with systems 
other than the gyroid mesophase. However, it is about an order of
magnitude faster and the general principles underlying it should be
applicable to different systems as well.

\section{A general pattern recognition based approach for the detection of
defects}
The first approach is based on the regularity or periodicity of patterns
and was developed by Chetverikov and Hanbury in 2001
\cite{bib:chetverikov}. It is assumed that the defect-free pattern is
homogeneous and shows some periodicity. The algorithm searches for areas
which are significantly less regular (i.e. aperiodic) than the bulk of the
dataset by computing regularity features for a set of windows and
identifying defects as outliers in regularity feature space. The
regularity is quantified by computing the periodicity of the normalised
autocorrelation function in polar coordinates. In short, for every window
a regularity value is computed. If this value differs by more than a defined
threshold value from the median of all window regularity values, the
area is accordingly classified as a defect. For a more detailed description
of the algorithm see \cite{bib:chetverikov,bib:chetverikov2}.

\begin{figure}[h]
\begin{center}
\includegraphics[height=6cm]{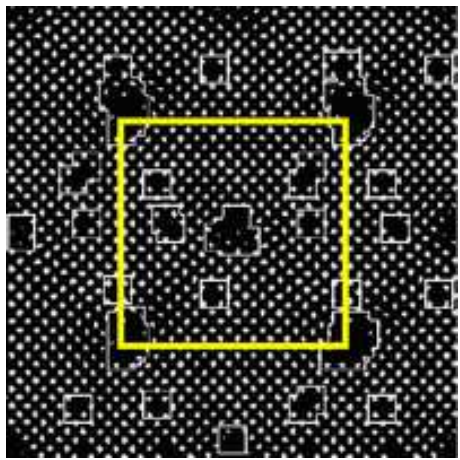}
\end{center}
\caption{Two-dimensional projection of the three-dimensional colour field
for a gyroid system of size 128$^3$ at $t$= 600000. The white frames
depict the areas detected by the pattern recognition algorithm of
Chetverikov and Hanbury \cite{bib:chetverikov}. In order to detect small
defects at the system's boundaries, we extend the dataset by mirroring
50\% of its corresponding content from the opposite side. The size of the
original dataset is depicted by the large box.}
\label{fig:colourstrucdef}
\end{figure}

We have found this pattern recognition algorithm to be very robust and
reliable in detecting defects. In figure \ref{fig:colourstrucdef} the
detected defects for an example dataset at $t$=600000 are depicted by the
white boxes. For good results, the regular pattern needs to occur multiple
times within an analysis window. A window size of 17x17 lattice sites has
been found to generate the best results. Defects located in the central
area of the crystal are very well detected, but the algorithm fails at the
boundaries. We overcome this problem by taking advantage of the fact that
our model has periodic boundary conditions, and extend the dataset by
mirroring 50\% of its corresponding content from the opposite side. In
this way we increase the number of analysis windows containing regular
patterns and the defects at the boundaries differ more substantially
from the surrounding regular pattern since they appear at their full size. The
pattern recognition algorithm is then able to detect boundary defects and,
depending on the maximum size of defects, the size of the additional
``padded'' regions can be adapted in order to limit the additional
computational costs. 

\begin{figure}[h]
\begin{center}
\includegraphics[height=5cm]{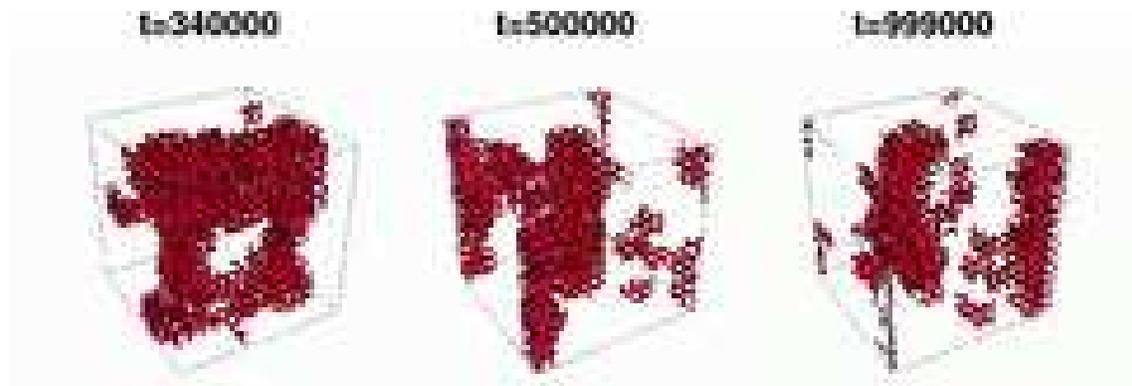}
\end{center}
\caption{Volume rendered visualization of the colour field at t=340000,
500000, 999000 from the evolving gyroid system. Only the defects are shown
as they have been isolated from the full datasets using the pattern
recognition algorithm.}
\label{fig:strucmaskreconst}
\end{figure}

By applying the pattern recognition algorithm to all individual projected
slabs of the dataset, we are able to reconstruct a three-dimensional
volume that only consists of areas which have been detected as defective. 
Figure \ref{fig:strucmaskreconst} shows reconstructed datasets at
$t$=340000, 500000 and 999000. Even at $t$=340000 a very large region of
the system has not yet formed a well defined gyroid phase. 160000
timesteps later, the main defects are pillar shaped ones at the centre and
at the corners of the visualised systems. Due to the periodic boundary
conditions, the corner defects are connected and should be regarded as a
single one. As can be seen from the analysis at $t$=999000, defects in the
gyroid mesophase are very stable in size as well as in their position. 

\section{A mesh generator as an alternative method for defect detection}
\begin{figure}[h]
\begin{center}
\includegraphics[width=135mm]{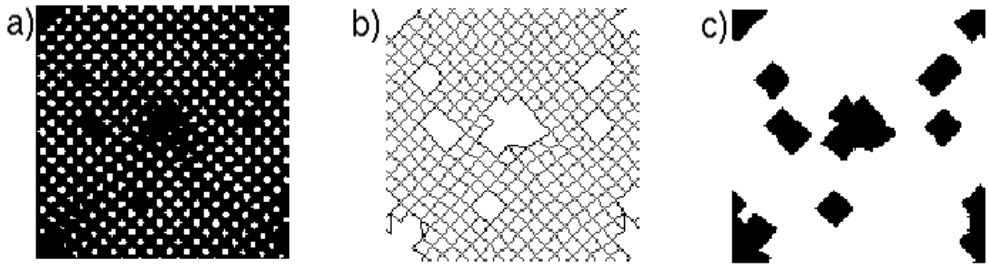}
\end{center}
\caption{
Defect regions (c) are extracted from a mesh representation (b) of the
two-dimensional projection (a) for $t$= 600000 shown in Figure \ref{fig:colourblack}.
}
\label{fig:compositemesh}
\end{figure}

The second approach presented in this paper also utilises the
two-dimensional slab projections and encapsulates knowledge about the
patterns produced by regular and defect regions (see figure
\ref{fig:compositemesh}). As a consequence, it is an order of magnitude
faster than the pattern recognition code.

\begin{figure}[h]
\begin{center}
\includegraphics[width=135mm]{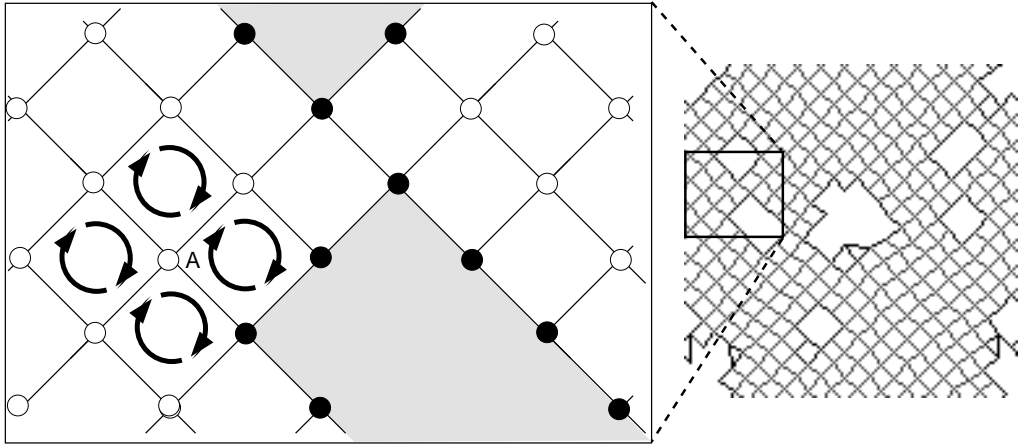}
\end{center}
\caption{Mesh nodes which lie within regular regions (A) are distinguished
by having four four-hop closed routes through their neighbouring nodes. Nodes
lacking this property (black) are categorised as defect boundary nodes.}
\label{fig:pathtracing}
\end{figure}

For each slab image, the centroid and area of each dot is computed and
dots below a certain threshold area are discarded. The four nearest
neighbours of each dot are determined and a connection mesh is generated
(Figure \ref{fig:compositemesh}(b)). Neighbours that lie more than one
gyroid unit cell width away from a dot are discounted.
Periodic boundary conditions are assumed when generating the mesh so that
defects at the edge of the system may be reliably detected.
Nodes which lie in defect-free regions of the lattice are distinguished by
having four four-hop closed routes through neighbouring nodes (denoted by
the arrows in figure \ref{fig:pathtracing}). Mesh nodes that describe the
perimeter of defect regions lack this property. A tree-searching algorithm
is used to search the data-structure representing the mesh and detect the
presence of closed loops.
The regions of regular mesh are discarded, leaving only mesh that
describes the perimeters of defect regions (emboldened regions in
\ref{fig:compositemesh}(b)).

A flood-fill algorithm is applied to a rasterised image of the defect
perimeters to locate distinct defect regions and isolate them from the
background (corresponding to defect-free regions). Thereafter, a mask
image as presented in Figure \ref{fig:compositemesh}(c) is generated.  

\begin{figure}[h]
\begin{center}
\includegraphics[height=5cm]{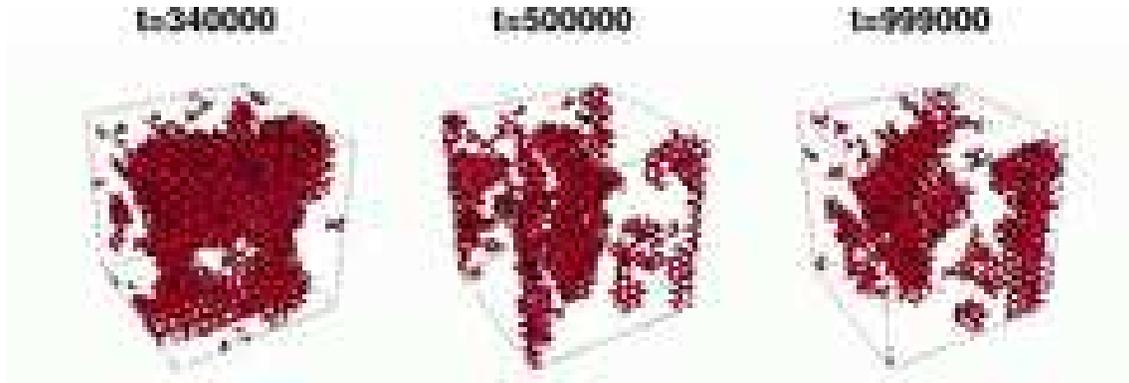}
\end{center}
\caption{Volume rendered visualization of the colour field at t=340000,
500000, 999000 from the evolving lattice-Boltzmann simulation. Only the
defects are shown as they have been separated from the full datasets using
the mesh generation algorithm.}
\label{fig:strucmeshreconst}
\end{figure}

As with the pattern recognition approach, this procedure is repeated for
each of the three Cartesian axes and the resultant mask images are
assembled to produce a three-dimensional defect mask which is then applied
to the original dataset. For comparison, figure
\ref{fig:strucmeshreconst} shows reconstructed visualizations of defective
areas for the same parameters and timesteps as figure
\ref{fig:strucmaskreconst}.

Since the algorithm defines the perimeter of a defect region by the
nearest mesh nodes, there is a tendency to over-estimate the boundary of
the defect region. However, this over-estimate is proportional to the
defect surface area and the maximum error of the detected defect volume
can be estimated as the volume of a one gyroid unit cell thick layer
surrounding all individual defects.

Since the dots in the two-dimensional projections are sections of tube-like structure that run through the gyroid rather than spatially localised entities, the division of the dataset into slabs is an essential step. For this reason, no attempt was made to develop a three-dimensional mesh generator. 

\section{Comparison of the detection algorithms}
\begin{figure}[h]
\begin{center}
\includegraphics[height=7cm]{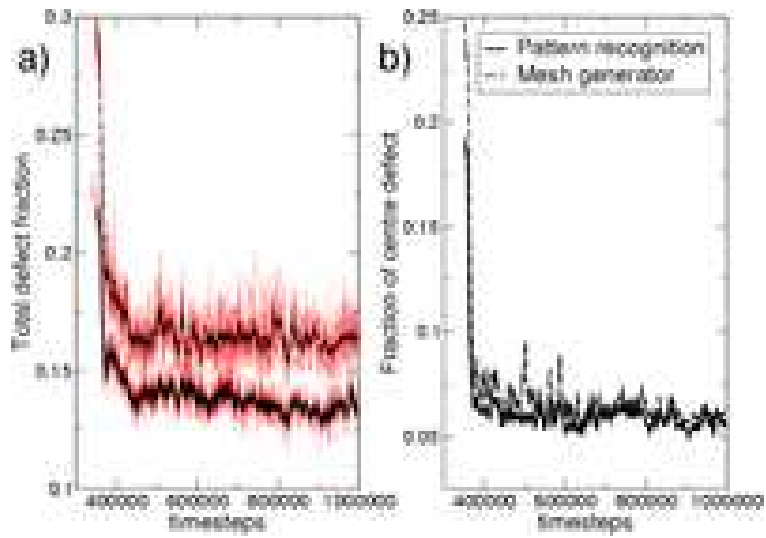}
\end{center}
\caption{Fraction of the total simulation volume contained in defects (a)
and volume fraction of the large central defect (b) detected by both
algorithms. The lighter noisy plots in (a) denote the original data from
the pattern recognition algorithm and the mesh generator. The solid and
dashed lines have been averaged over 4000 timesteps for better
visability.}
\label{fig:volumecomp}
\end{figure}

From the reconstructed datasets we are able to compute the volume fraction
of the simulation system that contains defects. This value is plotted in
figure \ref{fig:volumecomp}(a). The shades denote the original data, while
the solid and dashed lines are averages over 4000 timesteps. As expected,
the volume fraction detected by the mesh generator is larger than the area
detected by the pattern recognition algorithm because the mesh generation
algorithm's resolution is limited by the size of a unit cell. In addition,
the mesh generator detects very small and short-lived variations of the
dataset which occur due to small local variations of the gyroid structure
resulting in more noisy data for the volume fraction. The results of the
pattern recognition algorithm are noisy because the shape of the detected
regions is determined by the combination of overlapping squares
corresponding to the analysis windows. Small variations of the defect
shape can result in differently arranged overlapping windows and thus in
varying defect volume fractions. Improvement is possible by using a higher
resolution for the pattern recognition analysis, but at greater 
computational cost.
However, both methods show the same general behaviour. 

Figure
\ref{fig:volumecomp}(b) shows the averaged volume fraction contained in
the large defect in the centre of the system. We compute the total volume
of an individual defect by assuming that all detected areas that overlap
belong to the same defect. While for early simulation times, the mesh
generator's results fluctuate substantially more than the values obtained
from the pattern recognition algorithm, they eventually converge for
$t>$600000.

\begin{figure}[h]
\begin{center}
\includegraphics[height=6cm]{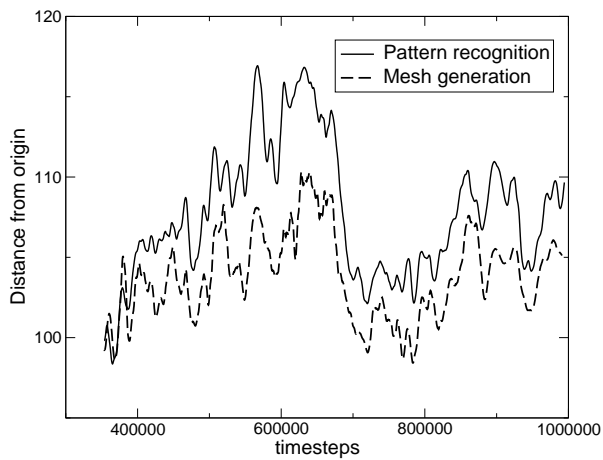}
\end{center}
\caption{Distance (in lattice sites) of the centre of the large defect in
the centre of the simulated 128$^3$ system from the origin. The data has
been obtained using the pattern recognition algorithm (solid line) and the
mesh generation algorithm (dashed line). The origin corresponds to the
left front corner in figures \ref{fig:strucmaskreconst} and
\ref{fig:strucmeshreconst}.}
\label{fig:distxyztot}
\end{figure}

An important feature of both algorithms is the possibility they provide to
track individual defects in time and so enable us to study their dynamics on the lattice.
As an example, we plot the distance of the centre of mass of the large
defect in the centre of the system in figure \ref{fig:distxyztot}. Both
methods generate the same general behaviour and differences in these plots
are caused by the slightly different volumes detected by both methods.
The origin corresponds to the lower front corner in figures
\ref{fig:strucmaskreconst} and \ref{fig:strucmeshreconst}.

Finally, we analyse the number of defects detected by both methods for
$t>$34000. As shown in table \ref{tab:numofdefs}, the pattern recognition
algorithm has a minimum detection rate of 4 defects which corresponds to a
value at an early simulation time ($t=$346200) and a maximum value of 34
($t=$578600). The mean is 20.17 and the standard deviation $\sigma$ is
3.48. All values are substantially smaller than the results obtained from
the mesh generator (see table \ref{tab:numofdefs}). This is because of the mesh
generator's ability to detect very small variations of the gyroid.
Furthermore, since the pattern recognition's resolution is limited due to
the rectangular shape of the analysis windows, the resulting detection
areas are not as flexible in shape as the ones from the mesh generator.
Thus, the mesh generator might detect multiple small defects which are
very close to each other while the pattern recognition algorithm would
detect those as a single defect. We can enhance the comparability of both
methods by applying a filter to the mesh generator and only taking defects
into account that have a lifetime of at least 1000 timesteps. For our
data, the new minimum drops to 7 which is only 3 defects more than
obtained using the pattern recognition algorithm. The resulting maximum is
36 and the mean becomes 22.76.

\begin{table}[h]
\begin{center}
\begin{tabular}{|l|c|c|c|c|}
\hline
\bf Algorithm & \bf Minimum & \bf Maximum & \bf Mean & $\mathbf{\sigma}$ \\
\hline
Pattern recognition & 4 & 34 & 20.17 & 3.48 \\
Mesh generator & 28 & 78 & 48.70 & 7.52 \\
Mesh generator (avg) & 7 & 36 & 22.76 & 3.80 \\
\hline
\end{tabular}

\end{center}

\caption{Statistics for the number of defects within the system obtained
from the pattern recognition algorithm, the mesh generator and the mesh
generator with added time averaging over 4000 timesteps.}
\label{tab:numofdefs}
\end{table}

In order to make the advantages of the two-dimensional projections
apparent, we have extended the pattern recognition algorithm to three
dimensions and applied it to the three-dimensional dataset directly. For
the case presented here, the computational effort needed to analyse a
single dataset is 2.7 times higher than the analysis of all individual
projections ($s$=8) in all three directions. Furthermore, the
three-dimensional approach does not allow a greater resolution than the
size of the analysis window which in our case is 17x17x17. The projection
based approach allows us to improve the resolution since the individual
two-dimensional windows are mapped to 17x17x8 volumes. By taking all three
cartesian directions into account, we are able to achieve an effective
resolution of 8x8x8. Furthermore, due to the well defined structures
produced by the projection algorithm, the detection rate of the pattern
recognition algorithm is higher than if one uses it on a non-processed
dataset.

\section{Conclusions}
We first briefly described the most spectacular liquid crystalline
mesophases which may arise in amphiphilic fluids, and our capability to
simulate these. The specific liquid crystalline mesophase of interest has
been the cubic gyroid phase. We then described two powerful algorithms
based on a pattern recognition algorithm and a mesh generation method to
detect and track defects in liquid crystals and applied them to simulation
data of a gyroid mesophase obtained during the TeraGyroid project
\cite{bib:teragyroid}. Both algorithms are superior to fully
three-dimensional approaches since they exploit basic properties of the
system to be analysed. The two-dimensional projection of slabs which have
the thickness of a crystal unit cell allows us to reduce the computational
analysis effort substantially. Since the pattern recognition algorithm was
not developed with gyroid mesophases in mind, it should be applicable to
many different regular structures. Additionally, it is more robust
in detecting defective areas than the mesh generation algorithm. However,
the latter is about ten times faster and thus saves a substantial amount
of CPU time if one has to analyse large amounts of simulation data. In
addition, we found that time averaging is efficient in filtering out short
time fluctuations or artefacts.

The methods described in this paper are most powerful if they are applied
in a combined fashion and it would be a natural extension to perform parts
of it during an ongoing simulation. For checking if a gyroid phase has
formed, observing the variation of the maxima of the projected structure
function of the order parameter is most efficient. Efficient parallel FFT
implementations are widely available and can be implemented within the
simulation code. If the variation of
the maxima of the projected structure function drops below a threshold
value, one should apply the mesh generation algorithm and track values
like the total defect volume or the number of defects. Since our
simulation code uses spatial domain decomposition, each CPU can generate
the two-dimensional projections of the order parameter individually
following which the mesh generation algorithm can be applied locally. The
computational effort for this analysis is negligible compared to the
actual simulation time, and moreover would allow the scientist to use
computational steering techniques
\cite{bib:chin-harting-jha,bib:brooke-coveney-harting,bib:harting-venturoli-coveney}
to monitor the state of the simulation while it is running.

The pattern recognition algorithm is less efficient than mesh
generation, but is the only choice if one is not limited to simulations of
gyroid mesophases. In the gyroid case, it is more efficient to use the
results from the mesh generator to select a smaller number of datasets for
post-processing using the pattern recognition algorithm since the
computational effort involved in the pattern recognition can be substantial. For
very large datasets, a promising approach is to determine regions of
interest within a single dataset using the mesh generation algorithm and
then to analyse subdomains of the system utilizing the pattern recognition
approach.

Articles are currently in preparation which make extensive use of the
detection and tracking algorithms described here in order to understand the
dynamics and properties of amphiphilic gyroid phases.

\section*{Acknowledgements}
We would like to thank N\'{e}lido Gonz\'{a}lez-Segredo for fruitful discussions and
Dmitry Chetverikov for providing the two-dimensional version of the ``The IPAN
Structural Defect Detector''. We would also like to thank him for his support
when we extended it to three dimensions. 

We are grateful to the U.K. Engineering and Physical Sciences
Research Council (EPSRC) for funding much of this research through RealityGrid
grant GR/R67699 and to EPSRC and the National Science Foundation (NSF) for funding the TeraGyroid project. 


\end{document}